\documentstyle{laa}     
\def\ut#1{\mathop{\vtop{\ialign{##\crcr
     $\hfil\displaystyle{#1}\hfil$\crcr\noalign
     {\kern1pt\nointerlineskip}\hbox{$\hfil\sim\hfil$}\crcr
     \noalign{\kern1pt}}}}}

\def\undersymbol#1#2{\mathop{\vtop{\ialign{##\crcr
     $\hfil\displaystyle{#2}\hfil$\crcr\noalign
     {\kern1pt\nointerlineskip}\hbox{$\hfil#1\hfil$}\crcr
     \noalign{\kern1pt}}}}}

\begin{document}
\thesaurus{13.07.1 ; 13.07.03}
\title{High-energy $\gamma$-ray emission from  GRBs}
\author{
    F. De Paolis   
    G. Ingrosso    
\and D. Orlando    
}
\offprints{G. Ingrosso}
\institute{
Dipartimento di Fisica, Universit\`a di Lecce, and INFN, Sezione di Lecce, 
Via Arnesano, CP 193, I-73100 Lecce, Italy }
\date{Received date; accepted date}
\maketitle
\begin{abstract}
GRBs are nowadays a rather well understood phenomenon in the soft (KeV-MeV)
$\gamma$-ray energy band, while only a few GRBs have been observed
at high photon energies ($E_{\gamma} \ut > 1$ GeV).
It is also widely recognized that GRBs accelerate protons to
relativistic energies and that dense media are often present nearby the 
sources. 
Within this framework and by further adopting Totani's suggestion that 
GRB events release an amount of energy 
$\sim 10^{54} \Delta \Omega$ erg,
we compute in detail the high-energy 
$\gamma$-ray flux from the decay of neutral pions produced through 
the interaction
of accelerate protons with nucleons in the surrounding medium.
We also take into account the local and intergalactic $\gamma$-ray absorption. 
The presence of magnetic fields around the GRB sources
causes the deflection of the accelerated protons and so a temporal spread 
of the produced high-energy $\gamma$-rays
with respect to the signal in the soft $\gamma$-ray band.
Moreover, we analyze the possibility to detect the $\gamma$-ray signal 
in the GeV-TeV energy range by the ARGO detector under construction in 
Tibet. 

\end{abstract}
\section{Introduction}

Gamma-ray bursts (GRBs) have intrigued observers and confounded theorists 
ever since their discovery over thirty years ago.
Up to now, the most meaningful contribution to the comprehension 
of the GRB physics has been obtained through the observations by
BATSE and EGRET  instruments on board of the CGRO satellite.
Since 1991, BATSE detects about one GRB per  day in the 
50 - 300 keV energy band (hereafter indicated as
{soft $\gamma$-ray} regime),
with fluences between $\sim 10^{-7} - 10^{-5}$ erg cm$^{-2}$ 
and durations ranging from $\simeq 10$ ms to $\simeq 10^3$ s
(Paciesas et al. \cite{paciesas}).

The accumulation of a substantial GRB population by BATSE shows that
they are isotropically distributed.
The simplest explanation for this fact is the GRB 
cosmological  origin, at variance with the the pre-CGRO view 
in which  GRBs were associated to neutron stars distributed in a thick 
galactic disk.

The GRB cosmological origin has been clamorously confirmed more recently 
by the Beppo-SAX observations (Costa et al. \cite{costa97}) 
of counterparts in the X-ray band of some GRBs.
Thanks to these observations, it has been possible not only a X-ray 
identification of GRBs, but also the determination with a better accuracy 
of the GRB source position, allowing the consequent tracking of optical and 
radio telescopes (Galama et al. \cite{galama}).

   \begin{table}
      \caption{For some bursts for which it has been possible to 
estimate the distance, we give their redshift $z$ and the energy 
${\cal E}^{iso}_{\gamma}$ radiated in the {soft $\gamma$-ray band}.
In the third column we indicate how the GRB redshift
has been measured: (a) from the redshift of the host galaxy; 
(b) from Lyman break of the optical afterglow; (c) from absorption lines 
in the optical afterglow.}
         \label{Table1}
\begin{center}
\begin{tabular}{|c|c|c|c|c|}
\hline
GRB        & $z$       & Note & ${\cal E}^{iso}_{\gamma}({\rm erg})$   & Ref.       \\
\hline
970228&0.695&a&$5.2\times 10^{51}$& Djorgovski \cite{djorgovski99a} \\
970508&0.835&c&$\sim 7\times 10^{51}$   &  Metzger \cite{met}  \\
971214&3.42 & a    & $\sim 3\times 10^{53}$ &Kulkarni \cite{kulkarni98} \\
980329&$\sim$5 & b    & $\ut > 10^{54}$     &Fruchter \cite{fruchter}\\
980613&1.0964&a&$5.2\times 10^{51}$ &Djorgovski \cite{djorgovski99b}      \\
980703&0.966&a&$ \ut >  10^{53}$    &Djorgovski \cite{djorgovski}      \\
990123&$\ut >1.6$&c&$ \ut > 3 \times 10^{54}$ &Kulkarni \cite{kulkarni99} \\
990510&1.62  &c &$     2.9 \times 10^{53}$ &Harrison \cite{harrison}   \\
991208&0.7055 &c &$   1.03  \times 10^{53}$ &Sagar \cite{sagar}  \\
991216&1.02 &a &$   6.7  \times 10^{53}$ &Sagar \cite{sagar}  \\
\hline
\end{tabular}
\end{center}
   \end{table}

In turn, redshift measurements (through optical observations) allow 
to evaluate the distance of some GRBs and therefore to derive the amount 
of energy they should have radiated in an isotropic explosion (henceforth
indicated as 
${\cal E}^{iso}_{\gamma}$) in the {soft $\gamma$-ray} energy band.

Currently, the most popular explanation for the GRB phenomenon is 
the fireball model (Rees \& M\'esz\'aros \cite{rees92}), i. e. the
dissipation of the kinetic energy of a relativistic
(with Lorentz factor $\Gamma \sim 10^2-10^3$) expanding blast wave 
in internal shocks (see Piran \cite{piran} for a recent review). 
Due to the short time scale variability of GRBs ($ \ut > 10^{-3}$ s), 
{soft $\gamma$-rays} are generally considered to be 
emitted by relativistic electrons, through synchrotron and 
inverse Compton mechanisms.
This model is particularly successful in explaining
the long wavelength afterglow behaviour observed for some GRBs.

In Table 1, for some bursts for which it has been possible to estimate the 
distance, we give their redshift $z$ and the inferred ${\cal E}^{iso}_{\gamma}$
values. Note that if GRBs are beamed sources as indicated by afterglow 
observations (Huang, Dai \& Lu \cite{huang}), the  emitted energy would be 
decreased by the factor $\Delta \Omega / 4 \pi$, where $\Delta \Omega$ is 
the beaming angle. As one can see in Table 1, the brightest burst 
(GRB 990123) has  ${\cal E}^{iso}_{\gamma} \ut > 3\times  10^{54}$ erg 
showing that in a GRB event up to one solar mass can be radiated in 
form of soft $\gamma$-rays.

It is also well known that GRBs are sources of {high-energy
$\gamma$-rays},
as it has been shown by the EGRET spark chamber, which
detected $\gamma$-rays in the energy band $\sim$ 30 MeV - 20 GeV, 
from a high percentage of bright BATSE bursts 
(Schneid et al. \cite{schneid1}, \cite{schneid2} and Hurley et al. 
\cite{hurley}).
These observations clearly show that the GRB phenomenon 
is not exclusively the domain of the {soft $\gamma$-ray} energy band.   

The natural question which raises is the production 
mechanism of the {high-energy $\gamma$-rays}
(with energy $E_{\gamma} \geq 1$ GeV) that, in principle, 
may be due to both electrons and protons.

However, if the electron emission spectra
which fit BATSE observations are extrapolated
in the high-energy domain, the obtained photon fluxes 
at Earth are well below the values detected by EGRET 
(Pilla \& Loeb \cite{pilla}). This fact manifestly 
implies that {high-energy $\gamma$-rays} from GRBs 
cannot be entirely accounted for 
by radiation emitted by accelerated electrons.

Indeed, as is usual for many high-energy  astrophysical sources 
(such as AGN, Seyfert galaxies, etc.), also for GRBs
it is expected that the most efficient mechanism to produce 
{high-energy $\gamma$-rays} is via $\pi^0$ decay of pions primarily 
produced through proton-nucleon ($pN$) interactions.
\footnote{
In principle, accelerated protons could radiate their energy through 
synchrotron emission, before strong interactions are in operation.
However, synchrotron emission is relevant only for 
protons with extremely high-energy $\sim 10^{20}$ eV 
(Vietri \cite{vietri}) and it is negligible for the production 
of $\sim 1$ GeV - 10 TeV photons, which are produced 
much more efficiently through $pN$ interactions.}

Within the framework of the fireball model for GRBs, it has been recently
argued that all GRB events could release roughly the same 
amount of energy 
${\cal E} \sim 10^{54} \Delta \Omega$ erg (Totani \cite{totani}).
\footnote{
We would like to mention that Kulkarni et al. (\cite{kulkarni00})
have suggested that afterglow observations may offer a robust method to 
evaluate the GRB energetics. In particular, from late time X-ray 
observations and assuming that the fraction 
of shock energy carried by electrons is close to equipartition, Freedman \&
Waxman (\cite {fw99}) estimate the GRB energy release ${\cal E}$
to to be in the range $3 \times 10^{51} - 3 \times 10^{53}$ erg,
which must be considered a more conservative range of (low) GRB energetics.
However, both the Totani (\cite{totani}) and the
Freedman \& Waxman (\cite {fw99}) estimates for ${\cal E}$
may still be considered consistent with observations, due to the 
uncertainties and precarious nature of GRB energy estimates
(see also Kulkarni et al. \cite{kulkarni00}).}

The total GRB energy is shared between electrons and protons which,
at least in the initial stage of internal shock generation,
carry a much larger amount of energy than electrons, by a factor 
$m_p/m_e \sim 2000$.

It is uncertain what fraction of the initial  energy is subsequently 
transferred by protons to electrons,
but the simplest Coulomb interaction cannot be very efficient
within the shortest GRB time-scale variability of $\simeq 10^{-3}$ s.
Another mechanism to transfer energy to electrons could be through 
proton synchrotron emission and subsequent annihilation 
$\gamma \gamma \rightarrow e^+e^-$, but also this process is generally 
inefficient owing to its quite long time-scale compared 
with the GRB variability.
\footnote{
It has been shown that the proton cooling time becomes comparable to 
the GRB duration only at the maximum proton energy ($\sim 10^{20}$ eV),
and, even in this case, the energy radiated in the GeV-TeV range 
is considerable only if the proton energy spectrum  is harder than the 
typical shock acceleration spectrum $J_p(E_p)\propto E_p^{-2.2}$
(Totani \cite{totani98a}, \cite{totani98b}).}

Indeed, according to the fireball model, the energy transfer from the proton 
into the electron component occurs via internal shock generation. The 
efficiency of this mechanism depends sensitively on the fireball Lorentz 
factor $\Gamma$ so that the spread of the $\Gamma$ values should account 
for the spread in the observed {soft $\gamma$-ray} emissivity 
${\cal E}_{\gamma}$ of the electronic component.

However, as one can see from Table 1 at least for the few GRBs for which 
it has been possible to estimate the distance, the energy 
${\cal E}^{iso}_{\gamma}$ 
is a minor fraction (between $\sim 10^{-4}$ and $\sim 0.1$) of 
the total GRB energetic assumed by Totani  (\cite{totani}).
So, it is likely that a substantial fraction of ${\cal E}$ remains 
in form of accelerated protons coming out from the GRB source with a total 
energy ${\cal E}_{p} \simeq 10^{54} \Delta \Omega$ erg. 
This is the reference value that we will adopt in our calculations.

It is by now widely recognised that GRBs accelerate protons at energies
even up $\sim10^{20}$ eV (Waxman \cite{waxman}, Vietri \cite{vietri95}, 
B\"ottcher \& Dermer \cite{bott}).
If it is so, these protons could in turn interact with
nucleons in the medium surrounding the GRB source, giving rise to $\pi^0$ and 
ultimately to a {high-energy $\gamma$-ray} flux on Earth. 
Clearly, a key requirement of this model is the existence of a dense enough 
cloud (which serves as a target for $pN$ interactions) surrounding
the GRB source or, alternatively, the existence of a cloud along the line of 
sight observer-burst and close enough to the source. 
Moreover, as we shall see in the following Sections, in order to have an 
observable high-energy $\gamma$-ray flux on Earth, the cloud number density 
$n_N$ has to be in the range $10^8 -10^{11}$ cm$^{-3}$.

At first sight, so high densities may seem unlikely. 
However, the presence of high density regions nearby GRB sources 
has been invoked in several cases. 
For example, Katz \cite{katz} has shown that a cloud with 
density $n_N \sim 2 \times 10^{11}$ cm$^{-3}$ and size $r_0 \sim 10^{15}$ 
cm is required to explain the delayed high-energy photons observed 
more than an hour following the GRB 940217. 
Moreover, Piro et al. (\cite{piro}) and Yoshida et al. (\cite{yoshida}) have 
reported an iron emission line in the X-ray afterglow spectrum of GRB 
970508 and of GRB 970828. The obtained line intensity requires a medium 
of density $n_N \simeq 10^{10}$ cm$^{-3}$, thickness 
$\Delta R \simeq 10^{14}$ cm and size $10^{16}$ cm
(Lazzati, Campana \& Ghisellini \cite{lazzati}).
The involved geometry of the target region could be 
either a cloud with a significant covering factor located off the line 
of sight burst-observer or a homogeneous spherical shell centered in 
the GRB progenitor.

Three main classes of models have been proposed for the origin of GRBs:
neutron star-neutron stars (NS-NS) mergers (Paczynski \cite{paczynski86}
and Eichler et al. \cite{eichler}), ``hypernovae'' or failed type Ib supernovae 
(Woosley \cite{woosley} and Paczynski \cite{paczynski98}) and 
``supranovae'' (Vietri \& Stella \cite{vietristella}).
In the NS-NS model the target cloud could be associated to an accretion disk 
around the GRB source. 
In the hypernova case, the burst is caused by the evolution of a massive 
($\simeq 100~M_{\odot}$) star located in a dense 
molecular cloud, while in the supranova scenario, a shell-like
supernova remnant is naturally left over around the location of the GRB. 

In all the above mentioned scenarios, high-energy $\gamma$-ray production 
via $pN$ interactions is expected with an intensity roughly proportional 
to $n_N$. Moreover, a prediction of our model is that 
a long wavelength afterglow emission from GRBs
should be present only if the cloud covering factor is low enough.

A feature of the model in question is a delay in the arrival times of the 
{high-energy $\gamma$-rays} with respect to the {soft $\gamma$-rays} 
and a temporal spread of the signal,
as a consequence of the proton deflection due to the presence 
of magnetic fields in the region surrounding the GRB source. 
A lower limit to the magnetic field is surely the typical 
interstellar magnetic field $B_{ISM} \simeq 1~\mu$G.

The aim of the paper is to estimate in detail the flux on Earth
of $\gamma$-rays in the GeV-TeV energy range from cosmological GRBs and in 
particular the temporal structure of the signal. 
In Sections 2 and 3 
we present the model we use to calculate the source function for 
$\gamma$-ray production via $pN$ interactions. In Section 4 we
present our results for the $\gamma$-ray flux on Earth, taking into account 
the intergalactic $\gamma$-ray absorption.
In Section 5 we briefly discuss the ground-based {high-energy 
$\gamma$-ray} experiments in connection with GRB observations and give
the model parameter range for which the GRB high-energy detection 
may be possible by the ARGO-YBJ experiment
(Abbrescia \cite{proposal}, Bacci \cite{addendum}).
Finally, our conclusions are presented in Section 6.

\section {The Model}

As previously stated, we assume that the GRB source
emits an amount of energy ${\cal E}_p \sim 10^{54} \Delta \Omega$ erg
in the form of relativistic protons,
released during a time $\Delta t\sim 1$ s, at a distance 
$r_0 \sim 10^{16}$ cm from the central GRB source
\footnote{
We have verified that our results do not sensitively depend on the
adopted values of $\Delta t$ and $r_0$.}.
We further assume that there exists a dense enough cloud near/around the GRB 
source so that accelerated protons subsequently interact with nucleons 
in the cloud giving rise to $\pi^0$ and ultimately to 
{high-energy $\gamma$-rays}.

An essential ingredient to calculate the $\gamma$-ray production rate
is the proton flux $J_p(E_p,r)$ which is a function of both the
proton energy and the radial coordinate $r$. 
As implied by Fermi type shock acceleration models
(see e.g. Gaisser \cite{gaisser}, Berezinskii et al. \cite{berez}),
$J_p(E_p,r)$ is assumed to follow a power-law energy spectrum. It moreover
depends on the distance $r$ from the central engine 
according to a function $f(r)$ to be later specified
\begin{equation}
\begin{array}{ll}
J_{p}(E_p,r) = C (E_p/{\rm GeV})^{-a_p} f(r) \\ \\
~~~~~~~~~~~~~~~~~~{\rm protons~cm^{-2}~s^{-1}~sr^{-1}~GeV^{-1}}~.
\end{array}
\label{eqno:pflux}
\end{equation}
Assuming $f(r_0) =1$, the constant $C$ is determined by 
the normalization condition which yields
\begin{equation}
C=\frac {1}{4 \pi r_0^2}\frac{1}{ {\cal I}(a_p)}
\frac{{\cal E}_p}{\Delta t}
\label{eqno:norma}
\end{equation}
where 
\begin{equation}
{\cal I}(a_p) = \int_{E_0}^{+\infty} (E_p/{\rm GeV})^{-a_p} E_p dE_p~,
\end{equation}
and $E_{0} = \Gamma m_{p} c^2$ is the minimum proton energy.

As far as the value of the energy spectral index $a_p$ is concerned, 
since protons are accelerated by shock waves moving with 
ultrarelativistic velocities and further
diffusive processes are negligible during 
the time before $pN$ interactions, we will adopt the relevant value  
$a_p \simeq  2.2$  (Bednarz \& Ostrowski \cite{bedn}, Vietri \cite{vietri2}).
\footnote{
In this respect we mention that the observed local CR spectrum has a power 
index $a_p \simeq  2.75$, softer than the previous one.
As it is well known, this is due to the fact that CRs diffuse throughout the 
Galaxy, scattering on the hynomogeneities of the magnetic field. In this 
diffusion process, high-energy particles rapidly loose energy and at the 
end of the diffusion process, the CR power index gets increased with respect 
to initial value.}

The radial profile $f(r)$ depends of course on the geometrical dilution 
factor which follows the $r^{-2}$ dependence.
Moreover, we take into account that protons 
emerging from the GRB source at the distance  $r_0$, 
subsequently contribute to the production of {high-energy $\gamma$-rays}
just in their first $pN$ interaction
(for simplicity we neglect interactions of secondary particles with
lower energy).
The probability that this first interaction occurs 
after a path of length $l$ is equal to
$\exp(-\sigma_{pN} n_{N} l)$,
where $\sigma_{pN}$ is the total $pN$ cross-section 
($\sigma_{pN}\sim 3 \times 10^{-26}$ cm$^{2}$, 
nearly independent of the proton energy) and 
$n_{N}$ is the nucleon number density 
of the cloud. 
Correspondingly, 
the average length travelled by protons in the surrounding 
medium results to be 
$\Delta R =(\sigma_{pN} n_N)^{-1} \simeq
      10^{-3}~{\rm pc}~(10^{10}~{\rm cm}^{-3}/n_N)$. 
The mass enclosed will be $M_{\rm cl}\sim 0.4~M_{\odot} (r_0/10^{15}~
{\rm cm})^2$ in a shell-like cloud geometry and 
$M_{\rm cl}\sim 0.2~M_{\odot} (10^{10}~{\rm cm}^{-3}/n_N)^2$ in the case of 
an intervening cloud.

An important point to stress is that {high-energy $\gamma$-rays} may in 
principle be re-absorbed if the path they have to travel inside the cloud 
(after being produced) is long enough. If the dominant absorption process 
is through $\gamma H$ interactions, the corresponding column density is 
$\simeq 3\times 10^{25}$ cm$^{-2}$ (Particle Data Group \cite{pdg}) or,
equivalently, an attenuation length $l_{\gamma H}\sim \Delta R$. 
In this case high-energy $\gamma$-rays should suffer attenuation. However, 
it is expected that clouds nearby the GRB source present a high ionization 
degree, so that the relevant absorption processes are $\gamma e$ and 
$\gamma p$ interactions, which have a much smaller cross-section implying  
a large $\gamma$ free-path before re-absorption.

Another effect we have to consider in defining the function $f(r)$
is the deflection of the proton trajectory before the 
interaction with target nucleons occurs. In fact, 
protons are deflected by the intervening magnetic field $B$
by an angle $\alpha(E_p) = l/R_L(E_p)$, 
where $R_L(E_p) = E_p /(eB)$ is the Larmor radius.
From simple geometrical considerations,
protons starting at $r=r_0$ and reaching the radial coordinate $r$
are deflected by an angle
\begin{equation}
\alpha \simeq \arcsin \left(\frac{r-r_0}{R_L(E_p)}\right)~,
\label{eqno:alfa}
\end{equation}
so that $l$ is given by
\begin{equation}
l(E_p,r) \simeq R_L(E_p) ~\arcsin \left(\frac{r-r_0}{R_L(E_p)} \right)
\label{eqno:l}
\end{equation}

Moreover, protons deflected by an angle $\alpha > \pi/2$, i.e. protons that 
have travelled for a length $l >\pi R_L(E_p)/2$, turn back towards 
the GRB source, so they presumably do not contribute to the production 
of observable {high-energy $\gamma$-rays}. 

On the basis of these assumptions, the radial profile takes the form:
\begin{equation}
\begin{array}{ll}
f(r) = \displaystyle{\left(\frac{r_0}{r}\right)^2}
~\exp[-\sigma_{pN} n_{N} l(E_p,r)]
          ~~~~~{\rm if} ~~l<R_L(E_p) \\ \\
f(r) = 0~~~~~~~~~~~~~~~~~~~~~~~~~~~~~~~~~~~~~~~~{\rm if} ~~l>R_L(E_p)~.
\label{eqno:fr}
\end{array}
\end{equation}

The effect of the presence of a magnetic field around the GRB source
is particularly relevant as regards the $\gamma$-ray production rate.
The ratio $\eta = R_L(E_p) \sigma_{pN} n_N$
between the scale length for proton deflection by magnetic fields
$R_L(E_p)$ and that for $pN$ interactions $(\sigma_{pN} n_N)^{-1}$,
sets the relative strength of the two effects, the deflection and the 
interaction.

In the case $\eta \ut < 1$, protons can be significantly deflected 
before interacting, so that most of the generated $\gamma$-rays will
not reach Earth. 
Now, since $R_L \propto E_p$, once $n_N$ and $B$ are given, 
the condition  $\eta \ut < 1$ means
\begin{equation}
E_p \ut < 10^3 ~{\rm GeV}
                \left (\frac{10^{10}~{\rm cm}^{-3}}{n_N} \right)
                \left (\frac{B           }      {1~\mu{\rm G}} \right)~,
\label{eqno:beta}
\end{equation}
where the magnetic field at $r>r_0$ has been taken of the order of 
magnitude of the interstellar magnetic field. 

\section{Source function}

As we discussed in the Introduction the relevant mechanism
for the {high-energy $\gamma$-ray} production with $E_{\gamma} > 1$ GeV
is the process
\begin{equation}
pN   \rightarrow ~ \pi^{0} + {\rm ~anything}~,
~~~~\pi^{0}~ \rightarrow ~\gamma \gamma~.
\label{eqno:interaction}
\end{equation}

Following Dermer \cite{dermer}, the production spectrum of 
{high-energy $\gamma$-rays} resulting from 
$pN$ collisions is given by 
\begin{equation}
\begin{array}{ll}
q_{\gamma}(E_{\gamma},r)=
16 \pi^2 n_{N} 
\displaystyle{\int_{E_{\pi}^{min}}^{+\infty}} dE_{\pi}
\displaystyle{\int_{E_{0}}^{+\infty}} dE_p~ J_p(E_p,r) \times \\ \\
\displaystyle{\int_{cos\theta_{max}}^{1}} dcos\theta~ E^{*} 
\displaystyle{ \frac{d^3 \sigma^{*}}{dp^{*3}} }
~~~~~~~~~{\rm \gamma~cm^{-3}~s^{-1}~GeV^{-1}}
\label{eqno:qgamma}
\end{array}
\end{equation}
where $\theta$ is the angle between the line of sight and the proton direction 
and asterisks refer to CM frame. 
Provided $-1\leq \cos\theta_{max} \leq 1$, one gets 
$\cos\theta_{max} = \Gamma E_{\pi} - E^{*}_{max}/(\beta \Gamma p_{\pi})$. 

According to Stephens-Badwar's model (Stephens \& Badwar \cite{sb}), 
the Lorentz invariant cross-section for $\pi^0$ production in $pN$ 
collisions inferred from high-energy experimental data 
can be parameterized in the form
\begin{equation}
E^* \frac{d^3 \sigma^*}{dp^{*3}}= A e^{-B p_{\perp}} 
(1-\tilde{x})^{C_1 - C_2 p_{\perp} + C_3 p_{\perp}^2}
\end{equation}
where
$\tilde{x}=(x_{\parallel}^{*2}+(4/s)(p_{\perp}^2+m_{\pi}^2))^{1/2}$,  
$x_{\parallel}^*=p_{\parallel}^*/p_{max}^*$, $s=2m_p(E_p+m_p)$ and
$A=140$, $B=5.43$, $C_1=6.1$, $C_2=3.3$, $C_3=0.6$.
   \begin{figure}[htbp]
   \vspace{14.5cm}
 \includegraphics{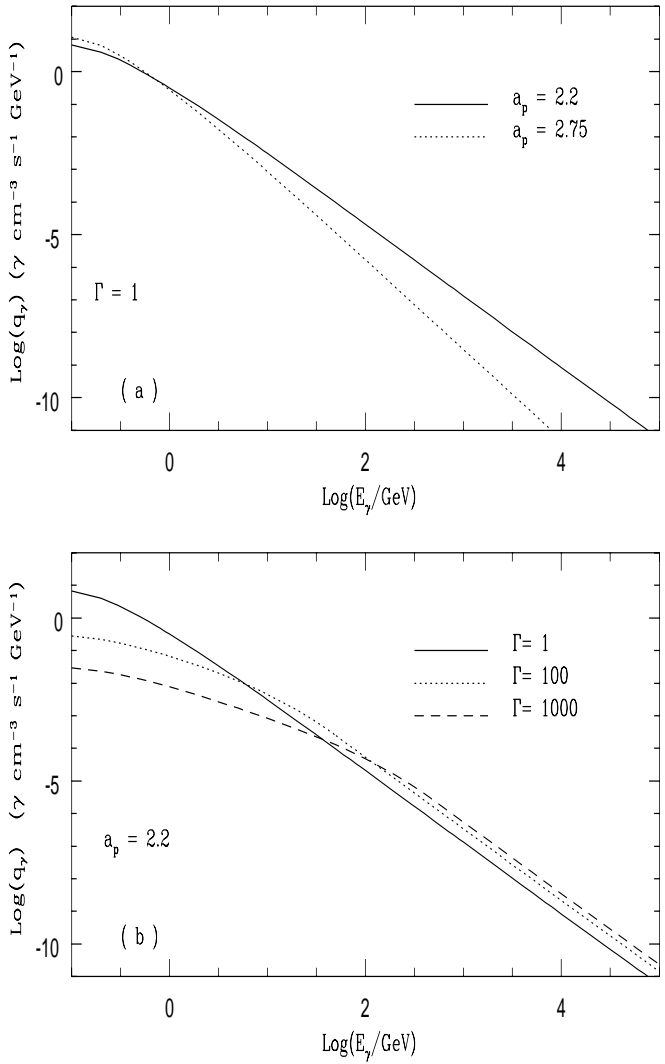} 
 \caption{Production spectra $q_{\gamma}(E_{\gamma})$ of 
  {high-energy $\gamma$-rays}
are given as a function of $E_{\gamma}$ for several values of $\Gamma$
and $a_p$, assuming $n_N=10^9$ cm$^{-3}$,
and $C=10^{23}/(4\pi{\cal I}(a_p))$ protons cm$^{-2}$ s$^{-1}$ sr$^{-1}$
GeV$^{-1}$.
In Fig. 1a, we set $\Gamma =1$ and give results
for $a_p=2.2$ (full line) and 2.75 (dotted line).
In Fig. 1b, for the relevant value $a_p =2.2$,
we see the effect on $q_{\gamma}(E_{\gamma})$ of increasing the values of
$\Gamma =1$, 100, 1000. }
            \label{Figure1}
   \end{figure}

Numerical values of the $\gamma$-ray source function 
in eq. (\ref{eqno:qgamma})
depend on the proton flux $J_p(E_p,r)$ at distance $r$, 
the Lorentz factor $\Gamma$ and the nucleon number density $n_N$.
The proton flux, in turn, depends on the constant $C$ and the
spectral index $a_p$.

In order to clarify the dependence of $q_{\gamma}(E_{\gamma},r)$ 
on the relevant parameters $a_p$ and $\Gamma$ we assume, as reference 
values, $n_N \simeq  10^9$ cm$^{-3}$ and 
$C \simeq 10^{23}/(4\pi{\cal I}(a_p))$ protons cm$^{-2}$ s$^{-1}$ sr$^{-1}$
GeV$^{-1}$ or, equivalently form eq. (\ref{eqno:norma}), 
$r_0  \simeq 10^{16}$ cm, 
${\cal E}_p \simeq  10^{54} \Delta \Omega$ erg and $\Delta t \simeq 1$ s. 

In Figs. 1a and 1b we give $q_{\gamma}(E_{\gamma})$ as a function of 
$E_{\gamma}$. 
In Fig. 1a $\Gamma$ is fixed to be one, while $a_p$ is 2.2 (continuous line) 
and 2.75 (dotted line). In Fig. 1b we set $a_p=2.2$, while $\Gamma$ values 
are 1, 100 and 1000.
As we can see in Fig. 1a, consideration of a harder proton spectrum increases
the {high-energy $\gamma$-ray} production rate (by a factor of $\sim 100$
at $E_{\gamma} \simeq 1$ TeV). 
Moreover, we find that the $\gamma$-ray spectrum roughly follows a power law 
behaviour $q_{\gamma}(E_{\gamma},r) \sim E_{\gamma}^{-a_{\gamma}}$, 
with power index $a_{\gamma}\simeq a_p$.  
In Fig. 1b, we see that an enhanced {high-energy $\gamma$-ray} 
production is also obtained by increasing the $\Gamma$ values.

Production spectra for different values of 
${\tilde C}$ and ${\tilde n_N}$ 
can be obtained from the results in Fig. 1 
by multiplying the quoted values by the factor
$ {\tilde C} ~( 4 \pi {\cal I}(a_p))/10^{23}) \times {\tilde n_N}/10^9$.
In particular, assuming ${\tilde n_N} = 1$ cm$^{-3}$ and
${\tilde C} = 2.1$ protons cm$^{-2}$ s$^{-1}$ sr$^{-1}$ GeV$^{-1}$
we can verify that our calculations reproduce, within a few percent,
the results on the $\gamma$-ray spectrum produced by the local 
cosmic-ray
interactions with the interstellar medium (Mori \cite{mori}).

\section {Gamma-ray flux on Earth}

In order to calculate the differential $\gamma$-ray luminosity 
$Q_{\gamma}(E_{\gamma}; r_0,R)$, 
we have to integrate the source 
function $q_{\gamma}(E_{\gamma},r)$ in eq. (\ref{eqno:qgamma}) over the 
whole volume of the region in which the $pN$ interactions occur, 
i.e. the shell between the radii $r_0$ and 
$R\simeq r_0 + 1/(\sigma_{pN} n_{N})$:
\begin{equation}
Q_{\gamma}(E_{\gamma}; r_0,R) =
\displaystyle{ \int_{r_0}^{R} }  q_{\gamma}(E_{\gamma},r) r^2 dr 
~~~{\rm \gamma~s^{-1}~GeV^{-1}}~.
\label{eqno:lum1}
\end{equation}
Note that here, as stated at the end of Section 2, we have
neglected re-absorption of the $\gamma$-rays in the cloud itself.

The explicit dependence of $Q_{\gamma}(E_{\gamma}; r_0,R)$ on 
the GRB parameters is obtained by using eq. (\ref{eqno:pflux}) 
in eq. (\ref{eqno:qgamma}), namely
\begin{equation}
\begin{array}{ll}
Q_{\gamma}(E_{\gamma}; r_0,R) =
\displaystyle{\frac{4 \pi n_N {\cal E}_p} {{\cal I}(a_p) \Delta t}}
\displaystyle{\int_{E_{\pi}^{min}}^{+\infty}} dE_{\pi}  
\displaystyle{\int_{E_p^{0}}^{+\infty}} E_p^{-a_p} dE_p ~ \times \\ \\
\displaystyle{\int_{r_0}^{R}}~ dr ~\exp[-\sigma_{pN} n_{N} l(E_p,r)] 
\displaystyle{\int_{cos\theta_{max}}^{1}} dcos\theta~ E^{*} 
\displaystyle{\frac{d^3 \sigma^{*}}{dp^{*3}}}~.
\end{array}
\label{eqno:lum}
\end{equation}

As we pointed out in Section 2, the presence of magnetic fields
around the GRB source affects the differential $\gamma$-ray luminosity 
in eq. (\ref{eqno:lum}) through the term $l(E_p,r)$.
   \begin{figure}[htbp]
   \vspace{7.5cm}
\includegraphics{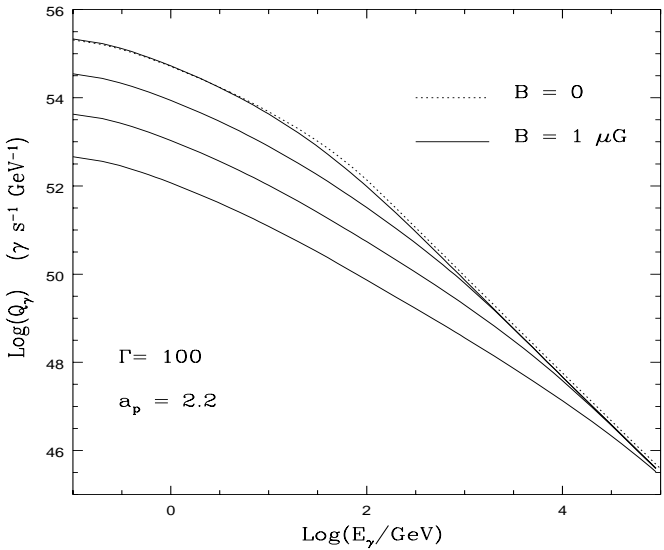} 
      \caption{The differential {high-energy $\gamma$-ray} 
       luminosity emitted by a GRB source via pN interactions is given 
       as a function of $E_{\gamma}$, for $B=0$ (dotted line) 
       and $B=1 ~\mu $ G (continuous lines).
       In the latter case, the cloud density values are
       $n_N = 10^8,~10^9,~10^{10},~10^{11}$ cm$^{-3}$,
       from the bottom to the top.
       Values of $Q_{\gamma}$ for different model parameters, 
       can be obtained by using the scaling relation
       $Q_{\gamma} ({\cal E}^{iso}_p/ 10^{55} ~{\rm erg})
       (\Delta t        /1 {\rm s})^{-1}$.}
       \label{Figure2}
   \end{figure}
In Fig. 2, adopting the same values for the GRB parameters as in Fig. 1 
and further assuming $a_p = 2.2$, $\Gamma = 100$
and $n_N$ in the range $10^8 - 10^{11}$ cm$^{-3}$,
we compare $Q_{\gamma}(E_{\gamma})$ obtained in the case
$B=0$ (in which protons follow straight trajectories) with the case
$B=1~\mu$G (in which protons of lower energies are deflected).
As one can see, in the case $B=0$, one obtains the same curve
for any value of the cloud number density. 
This result is expected since, 
increasing the density, the scale length for the $pN$ interactions decreases, 
but the projected column density $R n_N$ remains constant.

From Fig. 2 one can see that the effect of the presence of 
the magnetic fields is to diminish the $\gamma$-ray production rate, 
in the sense that for a given value of $B$, $Q_{\gamma}(E_{\gamma}; r_0,R)$
decreases as $n_N$ decreases. In fact, low energy photons are mainly 
generated by $pN$ interactions of low energy protons
(generally with energies of one order of magnitude above the photon energy), 
which are more easily deflected by the intervening magnetic fields and 
therefore give rise to photons that cannot reach the observer.
There exists, however, a limiting density 
($\simeq 10^{11}$ cm$^{-3}$ in the case $B \simeq 1~\mu$G)
above which most of the protons interact before suffer substantial
deflection so that the differential $\gamma$-ray luminosity 
for $B=0$ is recovered.

Another important effect of the presence of magnetic fields
is a temporal spread of the emitted {high-energy $\gamma$-ray}
signal on Earth. To better analyze this point, 
we can assume a proton injection time $\Delta t \simeq 1$ s, 
which may be considered instantaneous
since it is much smaller than the average proton interaction time 
$(\sigma_{pN} n_N c)^{-1} \simeq 10^5~{\rm s}~(10^{10}~{\rm cm}^{-3}/n_N)$.
Accordingly, denoting by $t_0$ the instant at which photons arrive 
on Earth in the case $B=0$ (no proton deflection), 
the delay $t-t_0$ is a function of the length $l$ travelled by the
protons (before interacting) and of the deflection angle $\alpha$
given by eq. (\ref{eqno:alfa}). So we get
\begin{equation}
\begin{array}{ll}
t_i-t_0 
= \displaystyle{ \frac{R_L(E_p)}{c}
\left(\arcsin \frac{r_i-r_0}{R_L(E_p)} - \frac{r_i-r_0}{R_L(E_p)}\right)}
\end{array}
\label{eqno:delay}
\end{equation}
Vice versa, one can say that photons with time delay between $t_1$ and $t_2$ are
produced by protons of energy $E_p$ only if these protons
have interacted in the shell $r_1(E_p)-r_2(E_p)$. Here 
$r_i(E_p) \equiv r(t_i,E_p)$, and
$r_i$ and $t_i$ are related by eq. (\ref{eqno:delay}).

We note that the relation above in eq. (\ref{eqno:delay})
allows to estimate the time duration $T_{HE}$ of the high-energy GRB signal.
If $\eta \ll 1$, namely the proton deflection is substantial
before $pN$ interactions take place, one gets $T_{HE}\simeq R_L(E_p) c^{-1}
\simeq 10^3~{\rm s}~(E_p/{\rm GeV})$. 
In other words, if one observe photons with energy 
$E_{\gamma} \simeq 100$ GeV (which are mainly produced by protons 
with $E_p \simeq 1$ TeV) the expected GRB duration at that energy
is $\simeq 10$ days.
If protons do not suffer strong deflection before interacting,
i.e. $\eta \gg 1$, one obtains $T_{HE} \simeq R_L(E_p) c^{-1} \eta^{-3}$
This implies that the high-energy GRB 
duration may become shorter and then more advantageous in a observational
perspective.

As next, let us estimate the differential $\gamma$-ray fluence on Earth
$F_{\gamma}(E_{\gamma};t_i,t_j)$,  
during the time interval $t_i-t_j$.
To this aim, we calculate $Q_{\gamma}(E_{\gamma};t_i,t_j)$ relative to 
photons with arrival time between $t_i$ and $t_j$, by substituting 
$r_0$ and $R$ with $r_i$ and $r_j$ in eq. (\ref{eqno:lum}) respectively. 
So:
\begin{equation}
\begin{array}{ll}
F_{\gamma}(E_{\gamma};t_i,t_j) = 
\displaystyle{ \frac{ Q_{\gamma}(E_{\gamma}(1+z);t_i,t_j) \Delta t (1+z)}
{4 \pi D_L^2(z)} }
\\ \\
~~~~~~~~~~~~~~~~~~~~~~~~{\rm \gamma~cm^{-2}~GeV^{-1}}~,
\end{array}
\label{eqno:flux}
\end{equation}
where $\Delta t$ is the proton injection time and
\begin{equation}
D_L(z) \simeq \frac{c}{H_0} \left(z + \frac{1}{2}(1-q_0)z^2\right),
\end{equation}
is the GRB luminosity distance, with $H_0 \simeq 70$ Km s$^{-1}$ Mpc$^{-1}$ 
and $q_0 \simeq 0.15$. 
Clearly, the  differential $\gamma$-ray flux on Earth at the time $t_i$ 
can be evaluated by
$\Phi_{\gamma}(E_{\gamma};t_i) = F_{\gamma}(E_{\gamma};t_i,t_i+dt)/dt$.

We note that by substituting eq. (\ref{eqno:lum}) into eq. (\ref{eqno:flux}) 
the injection time $\Delta t$ cancels out, so that
it does not affect the $\gamma$-ray signal on Earth, provided 
the observational time is $> \Delta t$.

It is interesting and important in connection with the observability
at high-energy of GRB events, to study in more details the temporal 
behaviour of the {high-energy $\gamma$-ray} flux arriving on Earth. 
To this purpose, in analogy to the GRB {soft $\gamma$-ray} observations,
we define the time $T_{50}(E_{\gamma})$ needed to receive 
on Earth the 50\% of the total number of photons of energy $E_{\gamma}$.
   \begin{figure}[htbp]
   \vspace{7.5cm}
\includegraphics{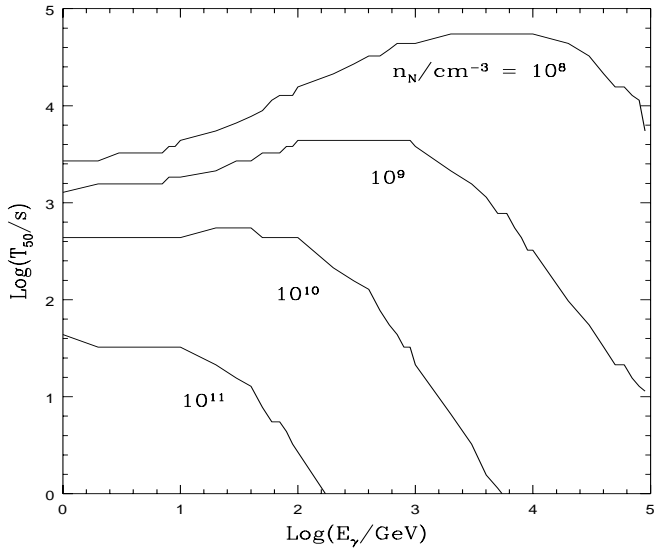} 
      \caption{The duration time $T_{50}$ is given as a function of 
$E_{\gamma}$ for selected values of the cloud density 
$n_N =~10^8,~10^{9},~10^{10},~10^{11}$ cm$^{-3}$.}
         \label{Figure3}
   \end{figure}
In Fig. 3 we show $T_{50}$ as a function of the photon energy, for different 
values of the cloud density $n_N$. As one can see, for each density value, 
there exist a limiting value of $E_{\gamma}$ above which $T_{50}$
decreases with increasing the photon energy.
Moreover, $T_{50}$  rapidly decreases as $n_N$ increases, for a given 
value of $E_{\gamma}$. 
From the same figure and the discussion above, it is clear 
that the higher is $n_N$, more easily the source is observable at high 
energy. This is essentially due to two reasons: the flux $\Phi_{\gamma}$ 
depends approximately linearly on $n_N$ and $T_{50}$ is shorter as 
$n_N$ increases. It is also obvious that, in order to accumulate a
significant amount of $\gamma$-ray photons, high-energy observations must 
last substantially longer if the cloud density $n_N$ is rather low.

As we stated in Section 1, it is 
nowadays widely recognized that GRBs are cosmological sources.
In this case {high-energy $\gamma$-rays} 
suffer attenuation (by electron-positron pair creation) 
due to their interaction with photons
of the intergalactic infrared radiation field (IIRF). 
This effect is taken into account through the absorption coefficient 
$\tau(E_{\gamma},z)$, which can be evaluated through 
the observational data on the IIRF and adopting theoretical models of the 
galactic spectral energy distribution of the IIRF from stars and dust 
reradiation  
(Stecker \& De Jager \cite{stecker1}, Salomon \& Stecker \cite{stecker2}).
   \begin{figure}[htbp]
   \vspace{7.5cm}
\includegraphics{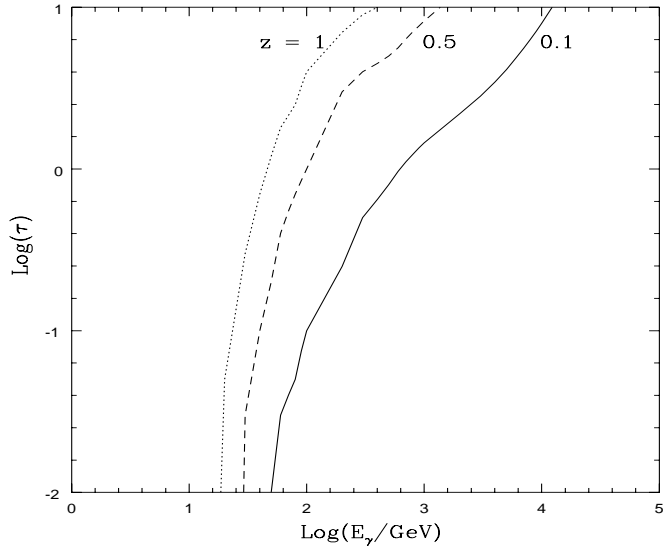} 
      \caption{The optical depth $\tau$ for high-energy $\gamma$-ray 
interactions with background low-energy photons is given as a function 
of the energy $E_{\gamma}$, for selected values of the source redshift
$z=0.1$, 0.5, 1.}
         \label{Figure4}
   \end{figure}
In Fig. 4, for illustrative purposes, we plot $\tau$ as a function 
of $E_{\gamma}$, for some values of the source redshift $z$. 
As one can see, for GRB sources at $z>0.5$, 
photons with energy $E_{\gamma} \ut > 100$ GeV are
heavily absorbed, while for closer GRBs with $z \ut < 0.1$
photons with $E_{\gamma} \ut < 1$ TeV survive absorption.
So, the differential $\gamma$-ray fluence observed at Earth will be 
\begin{equation}
F_{\gamma}^{\oplus}(E_{\gamma};t_i,t_j)=
F_{\gamma}(E_{\gamma};t_i,t_j) \exp[-\tau(E_{\gamma},z)]~.
\end{equation}
An analogous relation allows to calculate the differential 
$\gamma$-ray flux on Earth $\Phi_{\gamma}^{\oplus}(E_{\gamma};t)$. 
   \begin{figure}[htbp]
   \vspace{14.5cm}
\includegraphics{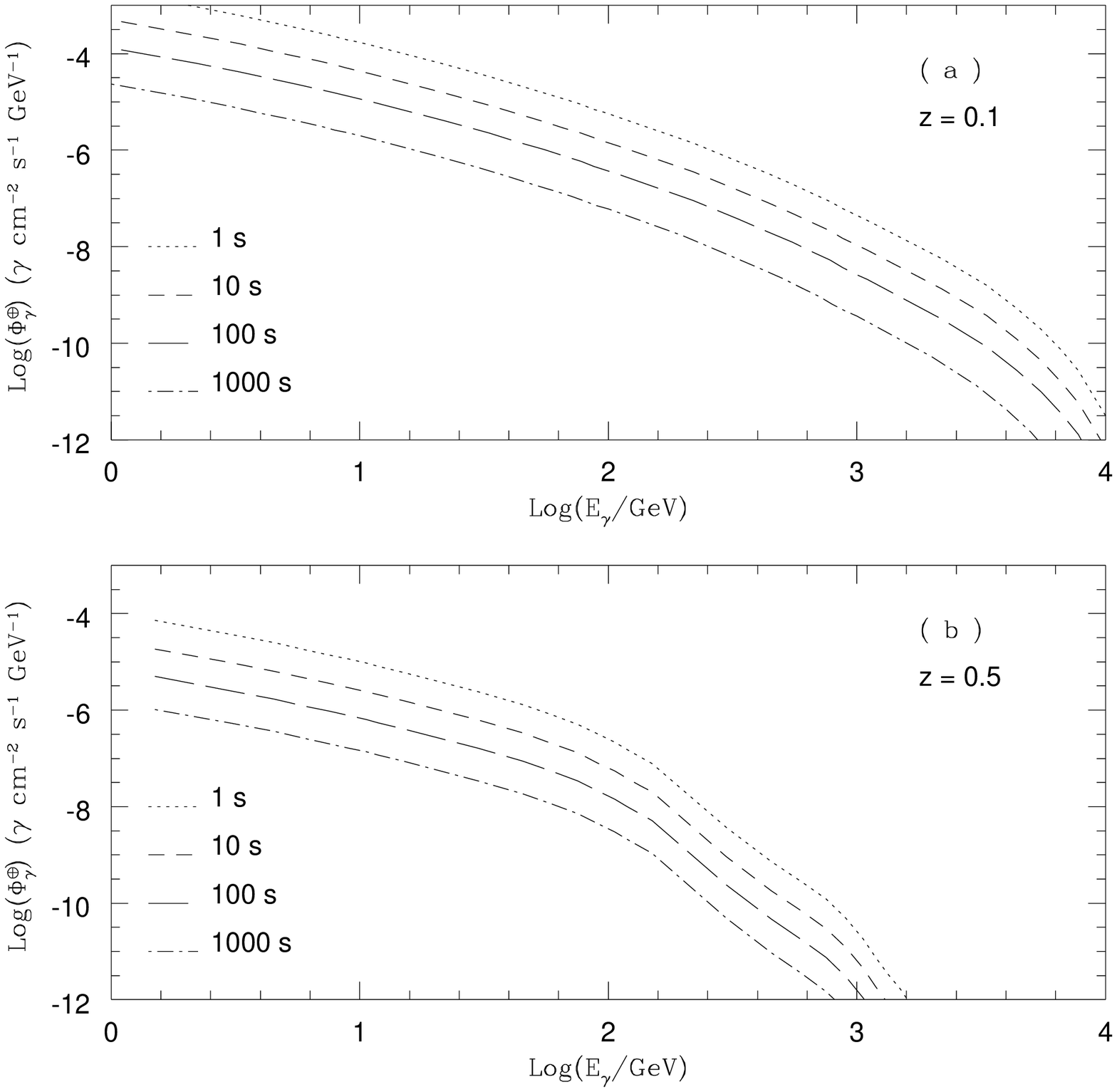} 
      \caption{Differential {high-energy $\gamma$-ray} fluxes on Earth 
are given at different time 
$t_i=1,~10,~100,~1000$ s. 
The GRB source redshift is fixed at the value $z = 0.1$ (a) and 
$z=0.5$ (b).}
         \label{Figure5}
   \end{figure}
In Figs. 5a and 5b, $\Phi_{\gamma}^{\oplus}$
at different arrival times is given as a function of the photon energy,
for GRB sources at $z=0.1$ and $z=0.5$, respectively.

In Fig. 6, for the same values of the source redshift $z$, the 
differential fluence on Earth is given in different
time intervals. As one can see in Figs. 5 and 6, due to the presence of 
magnetic fields,  GRBs may last much longer in the high-energy band 
than in the {soft $\gamma$-ray} band. 

   \begin{figure}[htbp]
   \vspace{14.5cm}
\includegraphics{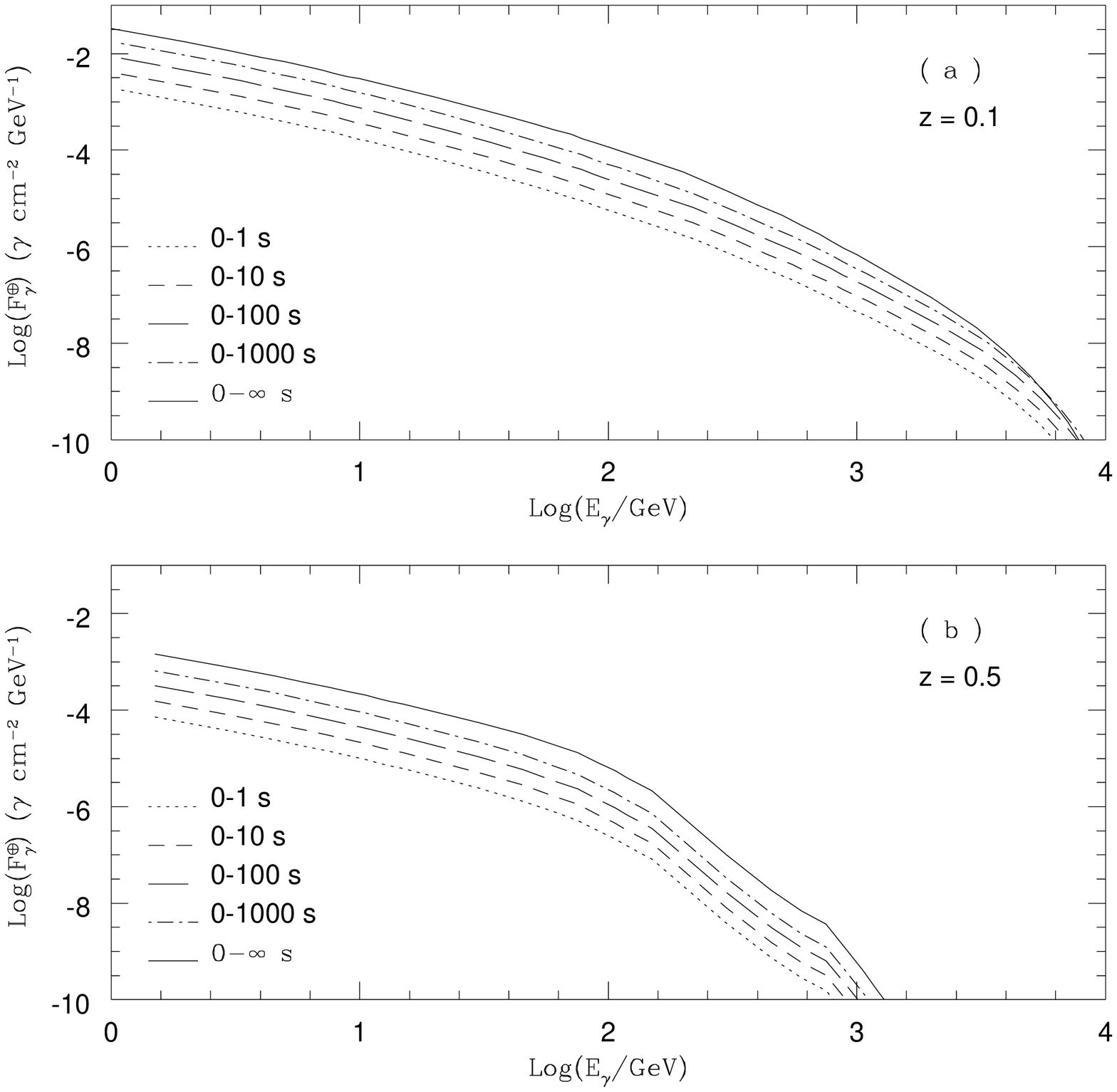} 
      \caption{Differential {high-energy $\gamma$-ray} fluences on Earth
      are given for different time intervals 
      $(t_i,t_j)=$ (0-1), (0-10), (0-100), (0-1000), (0,$\infty$) s. 
      As in Fig. 5a and 5b, the GRB redshift is  
      fixed to be $z = 0.1$ (a) and $z = 0.5$ (b).}
         \label{Figure6}
   \end{figure}

\section{Ground-based high-energy gamma-ray experiments}

Nowadays, the possibility to detect GRBs in the GeV-TeV energy range relies 
essentially on ground-based experiments
(see e.g. Vernetto \cite{vernetto1}). Less constrained in size 
than space born detectors, they can explore high-energy $\gamma$-rays  
detecting the secondary particles generated by their interaction  
in the atmosphere: 
Cherenkov photons in the case of Cherenkov telescopes 
as HEGRA AIROBICC (Padilla et al. \cite{padilla}),
Milagrito (McEnery et al. \cite{milagrito}) and
Whipple (Connaughton et al. \cite{whipple})
or $e^{\pm}$ in the case of extensive air shower array 
experiments as the Tibet air shower 
array (Amenomori et al. \cite{amenomori}) and the planned detector
ARGO (Abbrescia \cite{proposal}, Bacci \cite{addendum}).

The Tibet shower array and the HEGRA AIROBICC experiment  
have independently reported significant excesses of 10-20 TeV $\gamma$-rays 
in coincidence with a few GRBs.

In particular, among the 57 GRBs detected by BATSE 
(Meegan et al. \cite{meegan}) in the field of view of the
Tibet array, some of them show a significant excess of 
10 TeV $\gamma$-ray events with a timescale 
of $\sim 10$ s. The statistical significance was estimated to be about 6 
$\sigma$. 

There are four GRBs in the field of view of HEGRA array, and the HEGRA
group observed 11 $\gamma$-ray-like events above 20 TeV in a minute bin 
coincident with GRB 920925c in the GRANAT/WATCH GRB catalog 
(Sazonov et al. \cite{sazonov}), 
whereas only 0.93 events are expected as background (the statistical 
significance is therefore 5.4 $\sigma$). The chance probability, taking 
into account an appropriate trial factor, was estimated as 0.3\%. 
Seven out of the 11 $\gamma$-ray events are clustered within 22 s, 
suggesting that the GRB timescale is $\sim 10$ s.

More recently, the Cherenkov telescope Milagrito 
(McEnery et al. \cite{milagrito}) detected an excess of events 
within the BATSE error box of
GRB 970417a, with chance probability $2.8\times10^{-5}$.
Since all the 54 GRBs detected by BATSE and falling 
in the Milagrito field of view have been examined, 
the chance probability of observing an excess with this significance in any 
of these bursts is  $1.5\times10^{-3}$. 
Moreover, we note that the energy fluence of GRB 970417a 
may be explained with a source redshift $z \sim 0.7$ and an
isotropic total energy in the TeV range, 
${\cal E} _{TeV}^{iso} \ut > 10^{54}$ erg (Totani \cite{totani2000}).

From the above mentioned observations 
we can conclude that GRBs emit also in the GeV-TeV energy range,
although only a fraction of the GRBs detected by BATSE
have been observed at high-energy.
This is consistent with the fact that high-energy $\gamma$-rays
suffer serious attenuation 
due to the IIRF radiation, as discussed in Section 4.

In the framework of our model discussed in the previous Sections,
the possibility to detect high-energy GRB emission depends on the 
assumed model parameter values, in particular, on the GRB energetics
${\cal E}_p$, the cloud density $n_N$ and the magnetic field $B$. 

A systematic study of the parameter range allowing detection through several 
experiments (in operation or planned) will be presented elsewhere.
However, it is likely that the most suitable instrument to detect GRBs 
at energies $< 100$ GeV is the GLAST satellite, planned to be launched
by NASA in 2005 (Gehrels \& Michelson \cite{glast}).

Here,  for illustrative purposes, we refer to the 
ARGO-YBJ (Astrophysical Radiation with Ground-based Observatory at 
YangBaJing) detector, under construction at the YangBaJing High Altitude 
Cosmic Ray Laboratory (Tibet, China).
It is a cosmic-ray telescope optimized for the detection of small 
size air shower and will be devoted to set several issues in 
cosmic-ray and astroparticle physics including, in particular, GRB
physics in the energy range between $\simeq$ 50 GeV and $\simeq$10 TeV
(see Bacci et al. \cite{argo-grb} and Vernetto \cite{vernetto2}).

GRBs are detectable at high-energy if the number $N_{\gamma}$ of air showers 
due to $\gamma$-rays from one burst is significant larger than the 
fluctuation $\sigma = \sqrt{N_b}$ of the background shower number
$N_b$ due to cosmic-rays with arrival direction compatible with the burst 
position. 
So, in order to reduce the background and increase the detection sensitivity,
a good angular resolution is of major importance (see eqs. 
(\ref{eqno:signal}) and (\ref{eqno:fondo})). 

The angular resolution of the ARGO apparatus 
$\psi_{70}\simeq 5^0$ (corresponding
to $\epsilon = 0.7$) and the effective areas
$A_{eff}^{\gamma}(E_{\gamma})$ and $A_{eff}^{CR}(E_p)$,
for $\gamma$-rays and cosmic-rays, respectively,
have been numerically estimated through Montecarlo simulations 
by the ARGO collaboration (Surdo \cite{surdo})
\footnote{
In the energy range 
100 GeV - 5 TeV the effective areas increase from 
2 $\times 10^{7}$ cm$^2$ to 9 $\times 10^{8}$ cm$^2$, for $\gamma$-rays,
and from 5 $\times 10^{6}$ cm$^2$ to 8 $\times 10^{8}$ cm$^2$, 
for CRs.}.

The number of events detectable by ARGO in the energy range
$E_{min}-E_{max}$, during an observational time $T_{obs}$, is calculated
by the following relations:
\begin{equation}
N_{\gamma}=\epsilon 
\int^{E_{max}}_{E_{min}} 
F_{\gamma}(E_{\gamma};0,T_{obs}) 
A^{\gamma}_{eff}(E_{\gamma}) dE_{\gamma}
\label{eqno:signal}
\end{equation}
for the GRB signal and
\begin{equation}
N_b= 2\pi (1-\cos \psi_{70}) T_{obs}
\int^{E_{max}}_{E_{min}} 
\Phi^{CR}(E_p) A^{CR}_{eff}(E_p) dE_p
\label{eqno:fondo}
\end{equation}
for the cosmic-ray background.
Here, the cosmic-ray flux $\Phi^{CR}(E_p)$ 
(in units of cm$^{-2}$ s$^{-1}$ GeV$^{-1}$ sr$^{-1}$)
is given by (Honda et al. \cite{honda})
\begin{equation}
\Phi^{CR}(E_p)=
\left\{\begin{array}{l}
C_1~(E_p+m_p)^{-2.585}~e^{-\chi}~~~{\rm for}~E_p < 75 {\rm GeV} 
\nonumber \\
C_2~E_p^{-2.75} ~~~~~~~~~~~~~~~~~~~~~~{\rm for} ~E_p > 75 {\rm GeV}~,
\nonumber \end{array}
\right.
\end{equation}
where $\chi = -1.871/(0.97+\sqrt{E_p^2-m_p^2})$,
$C_1 = 1.085$ and  $C_2= 2.103$.

We take an energy range between $E_{min}= 10$ GeV and $E_{max}=5$ TeV, 
although we have verified that our results do not change substantially if 
1 GeV $\ut < E_{min} \ut < 50$ GeV (due to the effective area suppression)
and $E_{max} \ut > 5$ TeV (due to the signal and background flux 
suppression).
   \begin{figure}[htbp]
   \vspace{7.5cm}
\includegraphics{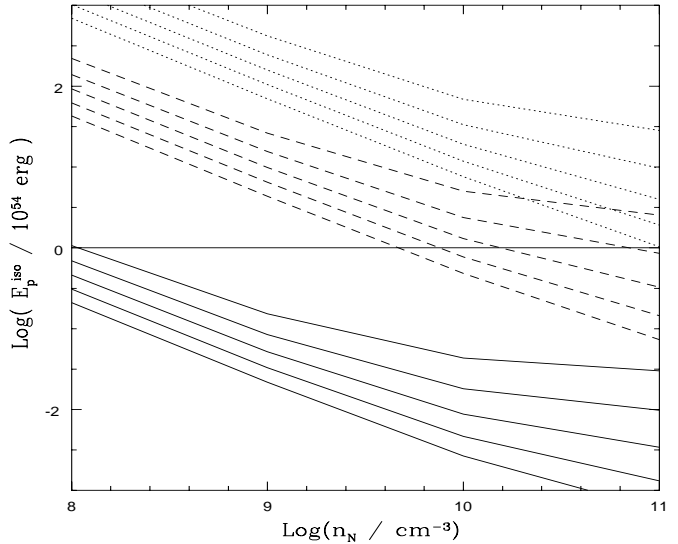} 
      \caption{
The isotropic GRB energetics ${\cal E}^{iso}_p$ (in units of $10^{54}$ erg)
is given as a function of $n_N$, for different intervals
of observational time 
($T_{obs} = 1,~10,~10^2,~10^3,~10^4$ s, from the bottom to the top) 
and GRB redshift $z=0.1$ 
(continuous lines), $z=0.5$ (dashed lines) and $z=1$ (dotted lines).
Each curve defines the model parameter region above which
GRBs should be detectable by ARGO apparatus, once the observational time 
and source redshift are given.}
         \label{Figure9}
   \end{figure}
In Fig. 7, assuming a detectability threshold condition
of 4 standard deviations (i.e. $N_{\gamma} \ge 4 \sqrt{N_b}$),
we give ${\cal E}^{iso}_p$ as a function of $n_N$, for different intervals
of observational time and for GRB redshift $z=0.1$ (continuous lines),
$z=0.5$ (dashed lines) and $z=1$ (dotted lines).
Each curve defines the model parameter region above which
a GRB should be detectable by ARGO apparatus, once the observational time 
and redshift are given.
The horizontal line in Fig. 7 marks a conservative (low) value for 
${\cal E}^{iso}_p = 10^{54}$ erg (corresponding to
${\cal E}_p \simeq 10^{53} \Delta \Omega$ erg). 
Nevertheless,  a GRB 
at $z=0.1$ produces a detectable high-energy $\gamma$-ray signal
for any cloud density value 
$10^8~{\rm cm}^{-3} \le n_N \le 10^{11}~{\rm cm}^{-3}$,
while if it lies at $z=0.5$ a density $n_N \ut > 5 \times 10^9$ cm$^{-3}$
is needed. Even a GRB at $z=1$ could be detectable provided 
$n_N \ut > 10^{11}$ cm$^{-3}$ 
and/or ${\cal E}^{iso}_p \ut > 10^{55}$ erg.

\section{Conclusions}

In this paper we adopted Totani's proposal that GRB sources release 
roughly the same amount of energy 
${\cal E} \simeq 10^{54} \Delta \Omega$ erg 
and that a substantial fraction of 
this energy is emitted in the form of accelerated protons coming out from the 
GRB source.

If there exists a dense enough cloud near or
around the GRB source, protons, interacting with the nucleons in the cloud,
give rise to pions and ultimately to a {high-energy $\gamma$-ray}
signal (in the GeV-TeV energy band) within the error box of a GRB 
detected in the KeV - MeV energy band.

Due to the presence of magnetic fields (with assumed intensity 
$B\simeq 1~\mu$G) around the source, i.e. within the proton propagation 
region before $pN$ interactions are in operation,
the {high-energy $\gamma$-ray} signal results to be delayed and 
spread out over a longer time with respect to the {soft $\gamma$-ray}
signal.

As it is shown in Fig. 3, the time duration of the $\gamma$-ray signal
in the GeV - TeV energy band crucially depends on the cloud density $n_N$
and lasts much longer with respect to the signal in the soft $\gamma$-ray 
regime.

By assuming that the accelerated proton injection time $\Delta t$ is 
much shorter than the proton interaction time, we 
have calculated the $\gamma$-ray flux on Earth at different times starting 
from the onset of the high-energy signal and the corresponding fluences 
for different integration times. These quantities depend on both the cloud 
density $n_N$ and redshift $z$ of the GRB source.
In particular, due to the $\gamma$-ray absorption from IIFR background 
radiation from stars and dust, the flux (and fluence) of the high-energy 
signal is largely suppressed for source distances $z>1$.

In Section 5 we have analyzed the possibility to detect 
a GRB in the high-energy band referring, in particular,
to the performances of the ARGO 
detector under construction in Tibet. 
As it can be seen from Fig. 7, there exists
a wide range of values for the relevant model parameters 
${\cal E}_p$ and $n_N$ which might make a GRB detectable in the
high-energy band.

As far as the GRB occurrence rate is concerned,
we rely on the analysis of the GRB space density done
by Schmidt \cite{schmidt} who assumed a homogeneous sample 
of GRB derived from BATSE DISCLA data and 
a wide variety of broken power-law GRB luminosity functions.
As a result of this analysis, one obtains that about 
13 GRB/yr are expected to occur at redshift $z \leq 0.5$ and $\sim$ 
120 GRB/yr at $z \leq 1$.

However, we note that a GRB with $z \sim 0.1$ is an extremely rare event 
(less than one GRB/yr). Moreover, it is expected that only a fraction 
of the GRB events meets the conditions to be observable on Earth
since we do not claim that all bursts have 
${\cal E}_p = 10^{54} \Delta \Omega$ erg and that accelerated protons interact 
with a very dense medium 
(with $10^8 {\rm cm}^{-3} \leq n_N \leq 10^{11} {\rm cm}^{-3}$).

Finally, we note that, although the details of the target cloud geometry 
are not important as far as the high-energy $\gamma$-rays flux estimates,
the long-wavelength afterglow crucially depends on that geometry.
In particular, it is expected that a long-wavelength afterglow may accompany 
the high-energy $\gamma$-ray production only if the cloud covering factor is 
low enough.

\acknowledgements{
We would like to thank the ARGO Group and in particular
B. D'Ettorre, G. Mancarella, G. Marsella, A. Surdo and S. Vernetto
for useful discussions.}

\end{document}